\documentclass[review]{elsarticle}
\usepackage{amsmath}
\usepackage{amssymb}
\usepackage{amsthm}

\begin{document}

\begin{frontmatter}

\title{Superconducting properties of a nonideal bipolaron gas}

\author[mymainaddress]{Victor Lakhno\corref{mycorrespondingauthor}}
\cortext[mycorrespondingauthor]{Corresponding author}
\ead{lak@impb.ru}

\address[mymainaddress]{Keldysh Institute of Applied Mathematics of RAS, 125047, Moscow, Russia}

\begin{abstract}
The properties of a Bose gas of translation-invariant (TI) bipolarons
 analogous to Cooper pairs are considered. As in the BCS theory,
the description of a TI-bipolaron gas is based on the electron-phonon
interaction and Froehlich Hamiltonian.  As distinct from the BCS theory,
when the correlation length greatly exceeds the mean distance between
the pairs, here we deal with the opposite case when the correlation
length is much less than the distance between the pairs.
We calculate the critical temperature of the transition of a TI-bipolaron
Bose-gas into the superconducting state, its energy, heat capacity and
heat of the transition. The results obtained are used to explain
the experiments on high-temperature superconductors.
Possible ways of raising the critical temperature
of high-temperature superconductors are discussed.
\end{abstract}

\begin{keyword}
high-temperature superconductors\sep translation-invariant (TI) bipolarons
\end{keyword}

\end{frontmatter}


\section{Introduction}

In this paper a translation-invariant (TI) bipolaron gas is considered as
a gas of Bose particles capable of forming a Bose condensate.
As is shown in \cite{1}-\cite{3}, a TI bipolaron is a state of two electrons coupled
by electron-phonon interaction (EPI) which remains delocalized for any value
of the EPI constant.  According to \cite{1}-\cite{3}, this state of electrons
is energetically more advantageous than a self-trapped localized
two-electron state or a Pekar bipolaron.

Since a TI-bipolaron gas is a gas of charged bosons, its Bose condensate
corresponds to a superconducting (SC) state. This fact is of interest
since nowadays numerous experimental and theoretical papers on high-temperature
superconductivity (HTSC) rely on the idea of bipolarons \cite{4}-\cite{8}.

It is important that the TI-bipolaron theory relies on the same initial
Froehlich Hamiltonian that the BCS theory does \cite{9}.  The only difference
is that in the BCS theory electrons interact with acoustical phonons,
while in the TI-bipolaron theory - with optical ones, because the properties
of HTSC are better explained in terms of this interaction.
As for the difference in the theoretical approaches, the BCS theory
is based on the exclusion of phonon variables from the Hamiltonian and
investigation of the resultant Hamiltonian containing only electron variables.
In the TI-bipolaron gas theory, on the contrary, the electron variables
are excluded from the Hamiltonian which results in the Hamiltonian
depending only on phonon variables.  The spectrum of eigen values of such
a Hamiltonian determines the spectrum of bipolaron states excitations.
The latter is just used to describe the statistic properties of an ideal bipolaron gas.

This approach, however, raises some important questions. Being charged,
a TI-bipolaron gas cannot be ideal since there must be a Coulomb interaction
between the bipolarons. According to the theory of a nonideal gas,
consideration of an interaction between particles leads to qualitative changes
in the spectral properties of the gas. According to \cite{10}, even in the case
of a short-range interaction, consideration of the latter results in the emergence
of a gap in the gas spectrum which is absent in the ideal gas. Still greater
changes would be expected in the case of a long-range Coulomb interaction.
Discussion of these problems is just the aim of this paper.

The logical scheme  of our approach is as follows:
a) first we consider a special case of two electrons interacting with a phonon field.
This is the classical problem of a bipolaron \cite{11}.

b) then we deal with a many-electron problem which leads to the concept of Fermi liquid.
For this many-electron problem we analyze the case of two additional electrons occuring
over the Fermi surface (near it) coupled by the EPI (Cooper's problem) \cite{12}.

c) then we believe that almost all the electrons occuring in the [$E_F+ E_{pol},E_F$] energy layer,
where $E_F$-is the Fermi energy, $E_{pol}$- is the polaron energy, are in the TI-polaron state \cite{12}; accordingly
 all electrons are in the TI-bipolaron states in narrow energy layer $[E_F + E_{bp}/2 - \delta E, E_F + E_{bp}/2 + \delta E], \delta E \rightarrow 0$, where $E_{bp}$ is the energy of TI-bipolaron. Condensed bipolaron gas leads to the infinite density of electronic states in this layer.

d) bipolarons are considered to be charged bosons placed in the electron Fermi liquid
(polaron gas) which screens the interaction between the bipolarons; in this case the
problem is reduced to that of a nonideal charged Bose gas.

e) the spectrum obtained upon solving this problem is used to calculate the statistic
properties of a TI-bipolaron gas.

As was noted above, the superconductivity theory is developed as a theory of a bipolaron
Bose gas where the superconducting state is a bipolaron condensate. As distinct from an
ordinary ideal Bose gas where superfluidity is absent, an ideal TI-bipolaron gas
demonstrates this phenomenon. Hence, in the theory under consideration,
superconductivity is a superfluidity of an ideal bipolaron gas. In an ordinary Bose gas,
superfluidity is possible only if bosons interact with one another and usually
takes place only at low temperatures. Being superfluid, an ideal bipolaron gas
retains this property even in the case of its nonideality.

The considered theory of a bipolaron gas can be used as the basis for the theory
of high-temperature superconductivity. The latter, however, has some limitations,
if for no other reason than because it, like the BCS theory, deals with the case
of a continuous medium. It is important, however, that being delocalized,
a TI bipolaron does not 'feel' discreteness of the lattice, nor the presence
of impurities or defects, if the interaction with them is not strong.

If the interaction of electrons with the lattice is strong or, what is the same,
an electron occuring at an atom of the lattice weakly interacts with neighboring
atoms, then the TI bipolaron whose interaction is based on Froehlich Hamiltonian,
breaks.  In this case the state of the electrons is described by discrete models.
The most popular superconductivity theories developed within this approach
are the model of resonating valence bonds (RVB) and t-J model \cite{13}.
They are based on Hubbard Hamiltonian and its generalization – Hubbard-Holstein Hamiltonian.
In the continuum approximation the latter is transformed to Froehlich Hamiltonian
which is basic for the TI-bipolaron theory.

The problems of developing the theory of superconductivity as a theory of Bose condensation
with the use of discrete models can be illustrated by the use of the small-radius
bipolaron theory for description of HTSC \cite{4}, \cite{5}. This theory relies on the idea that
 a stable bound bipolaron state is formed at one site of the lattice.
The small-radius bipolarons thus formed are considered as a gas of charged bosons.
However, really this approach can hardly be applied to HTSC since, on the one hand,
for small-radius bipolarons to be formed the electron-phonon interaction
(EPI) constant should be large, but on the other hand, for high value of Bose
condensation temperature we need a small mass, i.e. a small EPI constant \cite{14}-\cite{19}.
It is clear that the HTSC theory based on the concept of a small-radius bipolaron
which uses any other (non-phonon) mechanisms of interaction would encouneter the same problems.

The theory developed below will be shown to be free of this drawback.

The paper is arranged as follows. In \S2 we consider a problem of a nonideal
gas of TI-bipolarons which is reduced to the well-known problem of a charged Bose gas.
It is shown that due to screening in the system under consideration no gap
caused by interaction of Bose particles arises and actually the problem
is reduced to that of an ideal gas of TI-bipolarons having a gap in the spectrum.

In \S3 we deal with the thermodynamic properties of a TI-bipolaron gas.
For various values of the parameters which are the phonon frequencies,
we calculate the values of the critical temperatures of Bose condensation,
the heat of the transition to the condensate state, the heat capacity and
jumps in the heat capacity at the transition point.

 In \S4 the results are compared with experimental data.

In \S5 we sum up the results obtained and discuss possible ways of raising
the critical temperature of high-temperature superconductors.

\section{Nonideal TI-bipolaron gas}

To develop a theory of a nonideal TI-bipolaron gas we should know the spectrum
 of the states of an individual TI bipolaron in a polar medium. This problem
 was thoroughly discussed in \cite{11}. As is shown in \cite{12},  this spectrum of
the states will be the same as that of TI bipolarons arising in the vicinity
of the Fermi surface. Hence TI bipolarons in the [$E_F+E_{bp}/2,E_F$] layer can be considered
as a TI-bipolaron Bose gas occurring in the polaron gas \cite{20}.
If we believe that TI bipolarons do not interact with one another,
then the gas can be considered to be ideal. Its properties will be fully determined
if the spectrum of an individual TI bipolaron is known.

When considering the theory of an ideal gas and the theory of superconductivity
on the basis of Bose particles of TI bipolarons, Coulomb interaction between
the electrons is taken into account only for electron pairs, i.e. only for the problem
of an individual bipolaron. Hamiltonian of such a system, according to \cite{11}, has the form:

\begin{eqnarray}
\label {eq.1}
H_0=\sum_k\epsilon _k\alpha ^+_k\alpha_k
\end{eqnarray}

\begin{eqnarray}
\label {eq.2}
\epsilon _k=E_{bp}\Delta _{k,0}+\left(\omega _0+E_{bp}+k^2/2M_e\right)\left(1-\Delta _{k,0}\right)
\end{eqnarray}
where $\alpha ^+_k$, $\alpha_k$, -are operators of the birth and annihilation of TI bipolarons;
$\epsilon _k$ is the spectrum of TI bipolarons obtained in \cite{11};
$E_{bp}$ is the energy of a TI bipolaron; $M_e=2m^*$, $m^*$ is the electron effective mass;
 $\omega _0=\omega _0(\vec{k})$ is the energy of an optical phonon;
$\Delta _{k,0}=1$  for $k=0$  and $\Delta _{k,0}=0$ for $k\neq0$.
Expression (1), (2) can be rewritten as:

\begin{eqnarray}
\label {eq.3}
H_0=E_{bp}\alpha ^+_0\alpha_0+{\sum_k } '\left(\omega _0+E_{bp}+k^2/2M_e\right)\alpha ^+_k\alpha_k
\end{eqnarray}
where the prime in the sum  in the right-hand side of (3) means that the term with $k=0$
 is lacking in the summation. Extraction of the term with $k=0$
corresponds to the formation of a Bose condensate, where:

\begin{eqnarray}
\label {eq.4}
\alpha_0=\sqrt{N_0}
\end{eqnarray}
$N_0$ is the number of TI bipolarons in the condensed state.
Thus, in the theory of an ideal TI bipolaron gas, the first term is merely $E_{bp}N_0$.

In developing the theory of a nonideal TI-bipolaron Bose gas we will proceed from the Hamiltonian:

\begin{eqnarray}
\label {eq.5}
H=E_{bp}N_0+{\sum_k}'\left(\omega_0+E_{bp}\right)\alpha ^+_k\alpha_k+{\sum_k}'t_k\alpha ^+_k\alpha_k+ \\ \nonumber
1/2V{\sum_{k,k',k''}}'V_k\alpha^+_{k''-k}\alpha^+_{k'+k}\alpha_{k''}\alpha_{k'},\ \ t_k=k^2/2M_e,
\end{eqnarray}
This is Hamiltonian $H_0$ (3) plus the term responsible for bipolaron interaction;
$V_k$ is the matrix element of the bipolaron interaction.
The last two terms in (5) exactly correspond to the Hamiltonian of a charged Bose-gas \cite{21}.
Following the standard procedure of the Bose condensate resolving, we get from (5):
\begin{eqnarray}
\label {eq.6}
H=E_{bp}N_0+{\sum_k}'\left(\omega_0+E_{bp}\right)\alpha ^+_k\alpha_k+\\ \nonumber
{\sum_k}'\left[\left(t_k+n_0V_k\right)\alpha ^+_k\alpha_k+
1/2n_0V_k\left(\alpha _k\alpha_{-k}+\alpha ^+_k\alpha^+_{-k}\right)\right],
\end{eqnarray}
where $n_0=N_0/V$ is the density of the particles in a Bose-condensate.

Then, using Bogolyubov transformation:

\begin{eqnarray}
\label {eq.7}
\alpha_k=u_kb_k-v_kb^+_{-k},\\ \nonumber
u_k=\left[\left(t_k+n_0V_k+\epsilon_k\right)/2\epsilon_k\right]^{1/2},\\ \nonumber
v_k=\left[\left(t_k+n_0V_k-\epsilon_k\right)/2\epsilon_k\right]^{1/2},
\epsilon_k=\left[2n_0V_kt_k+t^2_k\right]^{1/2}
\end{eqnarray}
we get the Hamiltonian:
\begin{eqnarray}
\label {eq.8}
H=E_{bp}N_0+U_0+{\sum_k}'\left(\omega_0+E_{bp}+\epsilon_k\right)b ^+_kb_k,\\\nonumber
U_0={\sum_k}'\left(\epsilon_k-t_k-n_0 V_k\right),
\end{eqnarray}
where $U_0$ is the ground state energy of a charged Bose gas with
no regard for its interaction with the crystal's polarization.
Hence, the spectrum of excitations of a nonideal TI-bipolaron gas has the form:
\begin{eqnarray}
\label {eq.9}
E_k=E_{bp}+u_0+\\\nonumber
\left(\omega_0(\vec{k})+\sqrt{k^4/4M_e^2+k^2V_k n_0/M_e}\right)\\\nonumber
\ \left(1-\Delta_{k,0}\right),
\end{eqnarray}
where $u_0=U_0/N$, $N$ is the total number of particles.
If we reckon the energy of excitations from the  ground state enrgy of
a bipolaron in a nonideal gas, on the assumption that
$\Delta_k=E_k-(E_{bp}+u_0)$, then $\Delta_k$ for $k\neq0$ will be:
\begin{eqnarray}
\label {eq.10}
\Delta_k=\omega_0(\vec{k})+\sqrt{k^4/4M_e^2+k^2V_k n_0/M_e}
\end{eqnarray}
The spectrum obtained suggests that a TI-bipolaron gas has a spectrum gap $\Delta_k$
 between the ground and the excited states, i.e. it is superfluid. Being charged,
such a gas will be superconducting. To determine the particular form of the spectrum
we should know the value of $V_k$. If we considered only a charged Bose gas
with positive homogeneous background induced by rigid ion backbone,
then the value of $V_k$ in (9) would be equal to $V_k=4\pi e^2_B/k^2$
 in the absence of screening. Accordingly the second term in the radical
expression in (9) would be equal to $\omega^2_p=4\pi n_0e^2_B/M_e$,
where $\omega_p$ is the plasma frequency of the boson gas,
$e_B$ is the boson charge (2e for a TI bipolaron).
 Actually, if we take account of screening, then $V_k$ will take the form of
$V_k=4\pi e^2/k^2\epsilon _B(k)$, where $\epsilon _B(k)$  is the dielectric permittivity
of a charged Bose gas which was calculated in \cite{22}, \cite{23}.
The expression obtained for $\epsilon _B(k)$  in  \cite{22}, \cite{23} is too lengthy
and is not given here. In the case of a TI-bipolaron Boson gas under consideration
this modification of $V_k$ is not sufficient. As was shown in \cite{12}, bipolarons
are just few charged particles in the system. Most of the charged particles occur
in the electron gas where the bipolarons reside. It is just the electron gas that
makes the main contribution into the screening of the interaction between the bipolarons.
To take account of this screening  $V_k$ should be expressed as
$V_k=4\pi e^2/k^2\epsilon _B(k)\epsilon _e(k)$,
where $\epsilon _e(k)$  is the dielectric permittivity of the electron gas.
Finally, if we consider the mobility of the ion backbone, $V_k$ will take the form
$V_k=4\pi e^2/k^2\epsilon _B(k)\epsilon _e(k)\epsilon _{\infty}\epsilon _0$,
where $\epsilon _{\infty}$, $\epsilon_0$  are the high-frequency and static dielectric permittivities.

As a result, $\Delta _k$  is written as:
\begin{eqnarray}
\label {eq.11}
\Delta_k=\omega_0(\vec{k})+k^2/2Me\sqrt{1+\chi(k)}
\end{eqnarray}

\begin{eqnarray}
\label {eq.12}
\chi(k)=\omega^2_p/k^4\epsilon_B(k)\epsilon_e(k)\epsilon_{\infty}\epsilon_0
\end{eqnarray}
To estimate the value of $\chi(k)$  in (11) let us consider the long-wave limit.
In this limit $\epsilon_e(k)$ has Thomas-Fermi form: $\epsilon_e(k)=1+\varkappa^2/k^2$,
where $\varkappa=0,815k_F(r_s/a_B)^{1/2}$, $a_B=\hbar/M_ee^2_B$, $r_s=(3/4\pi n_0)^{1/3}$,
 therefore, according to \cite{22}, \cite{23}, the value of $\epsilon_B(k)$ is equal to $\epsilon_B(k)=1+q^4_s/k^4$,
$q_s=\sqrt{2M_e\omega _p}$. Bearing in mind that in calculations of thermodynamic
quantities the main contribution is made the values of $k$: $k^2/2M_e\approx T$,
 (where $T$ is the temperature), the value of $\chi(k)$ will be estimated to be
$\chi\backsim T/\epsilon _F\epsilon _{\infty}\epsilon _0$, where $\epsilon _F$
 is the Fermi energy. Hence the spectrum of the screened TI-bipolaron gas differs
from the spectrum of an ideal TI-bipolaron gas (2) only slightly.

Notice that due to screening the value of the correlation energy $u_0$ in (10)
turns out to be much less than the energy calculated in \cite{21} without regard for screening,
and for real values of the parameters, much less than the bipolaron energy
$\left|E_{bp}\right|$. Notice also that in view of screening a TI-bipolaron gas
does not form the Wigner crystal for arbitrarily small bipolaron density.

\section{Statistical thermodynamics of a low-density TI-bipolaron gas}

As was shown in \S2, a nonideal Bose gas of TI bipolarons differs from an ideal one only slightly.

Let us consider an ideal Bose gas of TI bipolarons as a system of $N$ particles
occurring in a volume $V$. Let us write $N_0$ for the number of particles in
the lower one-particle state, and $N'$ for the number of particles in the higher states. Then:
\begin{equation}\label{13}
    N=\sum_{n=0,1,2,...}\bar{m}_{n}=\sum_{n}\frac{1}{e^{(E_{n}-\mu)/T}-1},
\end{equation}
\begin{eqnarray}\label{14}
    N=N_0+N',\ \ N_0=\frac{1}{e^{(E_{bp}-\mu)/T}-1}, \\ \nonumber
		N'=\sum_{n_i\neq0}\frac{1}{e^{(E_{n}-\mu)/T}-1}.
\end{eqnarray}
In this section we will consider $\omega_0$ as independent of $k$.

In the expression for $N'$ (14), we will perform integration over
quasicontinuous spectrum (instead of summation) and assume $\mu=E_{bp}$.
As a result from (13), (14) we get an equation for the critical
temperature of Bose condensation $T_c$:
\begin{equation}\label{15}
    C_{bp}=f_{\tilde{\omega}}\left(\tilde{T}_c\right),
\end{equation}
$$f_{\tilde{\omega}}\left(\tilde{T}_c\right)=\tilde{T}^{3/2}_cF_{3/2}\left(\tilde{\omega}/\tilde{T}_c\right),$$
$$F_{3/2}(\alpha)=\frac{2}{\sqrt{\pi}}\int^{\infty}_0\frac{x^{1/2}dx}{e^{x+\alpha}-1},$$
$$C_{bp}=\left(\frac{n^{2/3}2\pi\hbar^2}{M_e\omega^*}\right)^{3/2},$$
$$\tilde{\omega}=\frac{\omega_0}{\omega^*},\ \ \tilde{T}_c=\frac{T_c}{\omega^*},$$
where $n=N/V$. Relation of the notation $F_{3/2}$ with other notaions is given in the Appendix.

Fig.~\ref{fig1} shows a graphical solution of equation (15) for the values of the parameters
$M_e=2m^*=2m_0$, where  $m_0$ is the mass of a free electron in vacuum,
$\omega^*=5$ meV ($\approx$58K), $n=10^{21}$ cm$^{-3}$ and the values:
$\tilde{\omega}_1=0,2$; $\tilde{\omega}_2=1$; $\tilde{\omega}_3=2$;
$\tilde{\omega}_4=10$, $\tilde{\omega}_5=15$, $\tilde{\omega}_6=20$.

\begin{figure}
\begin{center}
\includegraphics[width=0.9\textwidth]{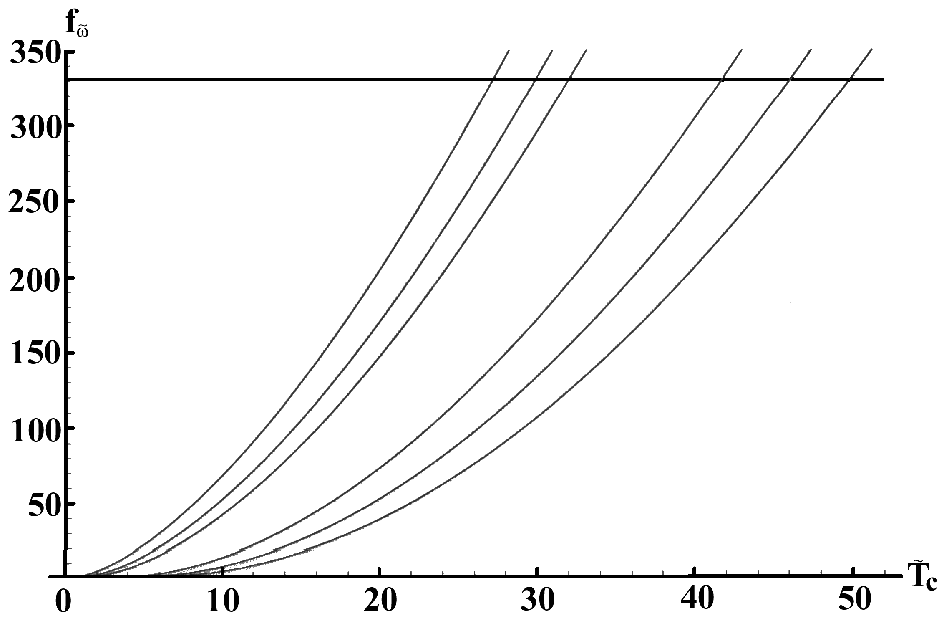}
\\
\includegraphics[width=0.9\textwidth]{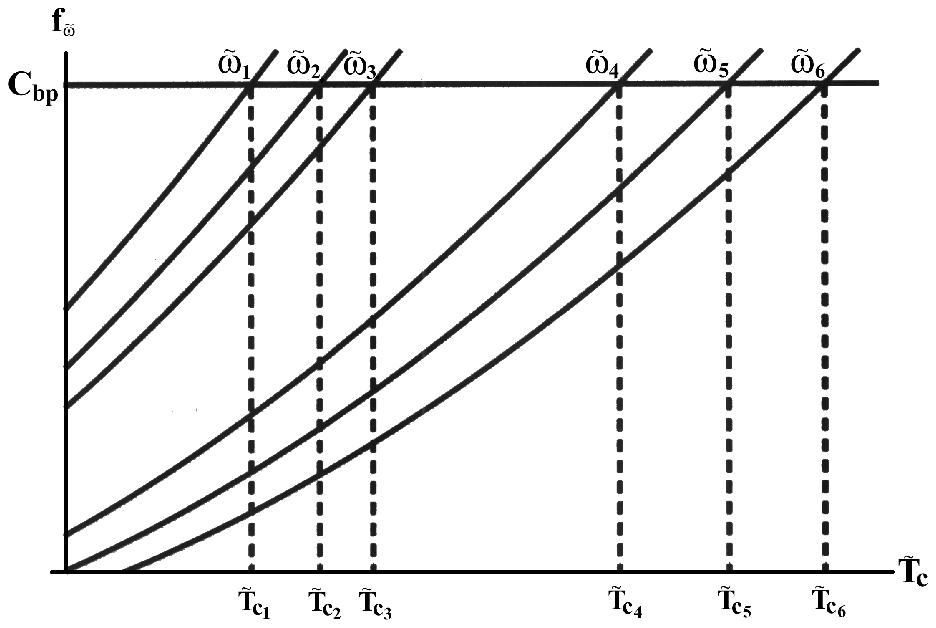}
\end{center}
\caption{Solutions of equation \eqref{15} with $C_{bp}=331,3$ and $\tilde{\omega}_i=\left\{0,2; 1; 2; 10; 15; 20\right\}$,
which correspond to
$\tilde{T}_{c_i}$: $\tilde{T}_{c_1}=27,3$; $\tilde{T}_{c_2}=30$;
$\tilde{T}_{c_3}=32$;
$\tilde{T}_{c_4}=42$; $\tilde{T}_{c_5}=46,2$; $\tilde{T}_{c_6}=50$.
\label{fig1}}
\end{figure}

\noindent It is evident from Fig.~\ref{fig1} that the critical temperature grows as the phonon
frequency $\omega_0$ increases. The relations of the critical temperatures
$T_{ci}/\omega_{0i}$ corresponding to the parameter values chosen are listed in Table~\ref{tab1}.

\begin{table}[h]
	\centering
		\begin{tabular}{|c|c|c|c|c|c|c|c|} \hline
		i & 0 & 1 & 2 & 3 & 4 & 5 & 6  \\ \hline
		$\tilde{\omega}_i$ & 0 & 0,2 & 1 & 2 & 10 & 15 & 20 \\ \hline
		$T_{ci}/\omega_{oi}$ & $\infty$ & 136,6 & 30 & 16 & 4,2 & 3 & 2,5 \\ \hline
		$q_i/T_{ci}$& 1,3 & 1,44 & 1,64 & 1,8 & 2,5 & 2,8 & 3 \\ \hline
		$-\Delta(\partial C_{v,i}/\partial\tilde{T})$ & 0,11 & 0,12 & 0,12 & 0,13 & 0,14 & 0,15 & 0,15 \\ \hline
		$C_{v,i}(T_c-0)$ & 1,9 & 2,16 & 2,46 & 2,7 & 3,74 & 4,2 & 1,6 \\ \hline
		$(C_s-C_n)/C_n$ & 0 & 0,16 & 0,36 & 0,52 & 1,23 & 1,53 & 1,8 \\ \hline
		$n_{bp_i}\cdot$cm$^{3}$ & 16$\cdot 10^{19}$ & 9,4$\cdot 10^{18}$ & 4,2$\cdot 10^{18}$ & 2,0$\cdot 10^{18}$ &
		1,2$\cdot 10^{17}$ & 5,2$\cdot 10^{14}$ & 2,3$\cdot 10^{13}$ \\ \hline
		\end{tabular}
\caption{Calculated characteristics of Bose-gas of TI-bipolarons with concentration $n=10^{21}$ cm$^{-3}$.
\label{tab1}}
\end{table}

Table~\ref{tab1} suggests that the critical temperature of a TI-bipolaron gas
is always higher than that of an ideal Bose gas (IBG).
It is also evident from Fig.~\ref{fig1} that an increase in the concentration of TI bipolarons
$n$ will lead to an increase in the critical temperature, while a gain
in the electron mass $m^*$ - to its decrease. For $\tilde{\omega}=0$,
the results go over into the well-known IBG limit. In particular,
(15) for $\tilde{\omega}=0$ yields the expression for the critical temperature of IBG:

\begin{equation}\label{16}
    T_c=3,31\hbar^2n^{2/3}/M_e
\end{equation}
It should be stressed however that (16) involves $M_e=2m^*$,
rather than the bipolaron mass. This resolves the problem of the low temperature
of condensation which arises both in the small-radius polaron theory and
in the large-radius polaron theory where expression (16) involves
the bipolaron mass ~\cite{24}-\cite{27}. Another important result
is that the critical temperature $T_c$ for the parameter values calculated
considerably exceeds the energy of the gap $\omega_0$.

\begin{figure}
\begin{center}
\includegraphics[width=0.9\textwidth]{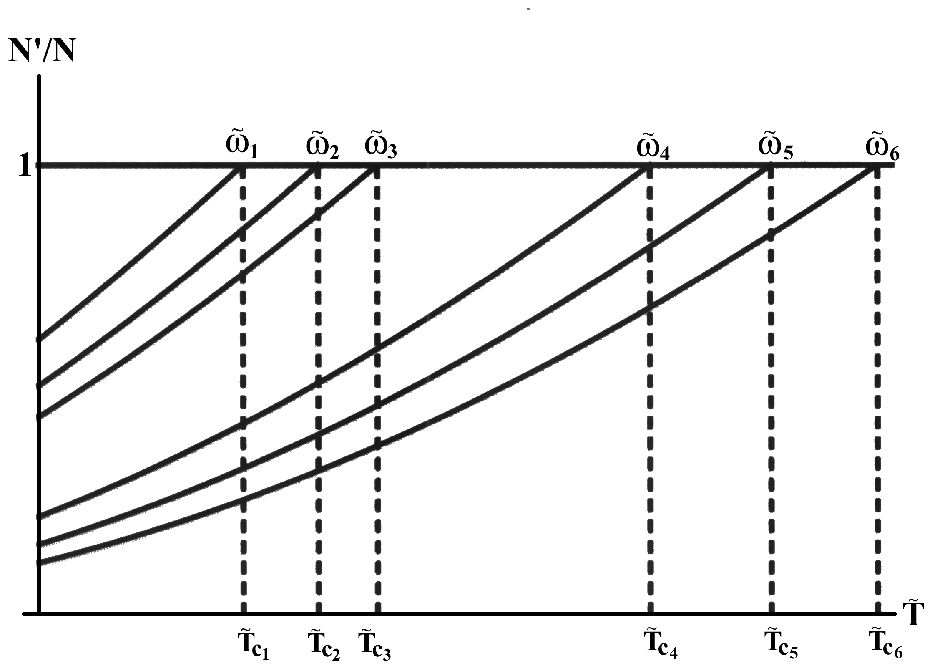}
\\
\includegraphics[width=0.9\textwidth]{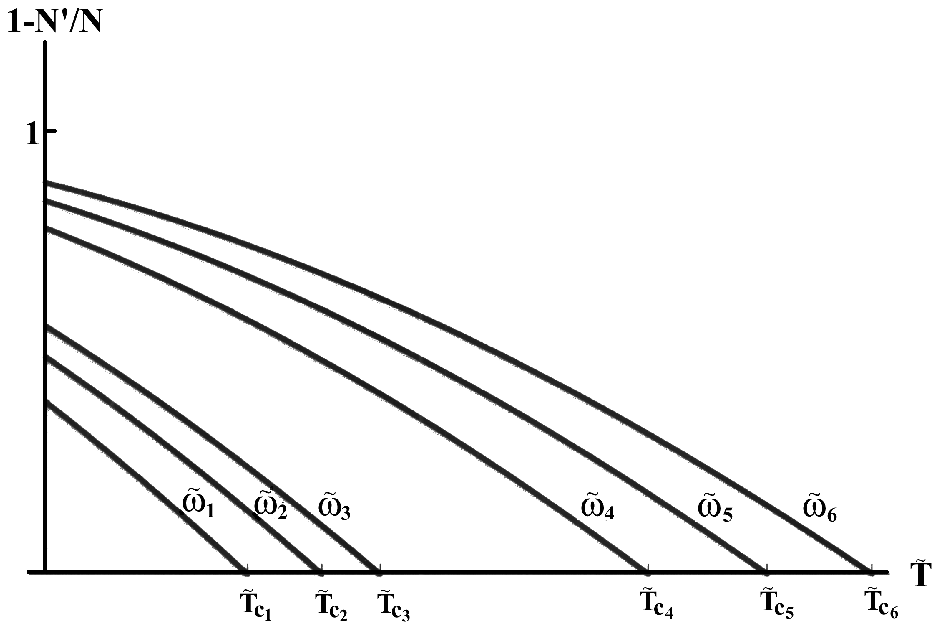}
\end{center}
\caption{Temperature dependencies of the relative number of supracondensate particles
$N'/N$ and the particles occurring in the condensate $N_0/N=1-N'/N$
for the parameter values $\tilde{\omega}_i$, given in Fig.~\ref{fig1}.
\label{fig2}}
\end{figure}

From (13), (14) it follows that:
\begin{equation}\label{17}
   \frac{ N'(\tilde{\omega})}{N}=\frac{\tilde{T}^{3/2}}{C_{bp}}
	F_{3/2}\left(\frac{\tilde{\omega}}{\tilde{T}}\right)
\end{equation}
\begin{equation}\label{18}
     \frac{ N_0(\tilde{\omega})}{N}=1-\frac{ N'(\tilde{\omega})}{N}
\end{equation}
Fig.~\ref{fig2} shows the temperature dependencies of the number of supracondensate particles $N'$
and the number of particles in the condensate $N_0$
for the above-listed parameter values $\tilde{\omega}_i$.

\noindent Fig.~\ref{fig2} suggests that, as would be expected, the number of particles
in the condensate grows as the gap increases $\omega_i$.

The energy of a TI-bipolaron gas $E$ is determined by the expression:
\begin{equation}\label{19}
    E=\sum_{n=0,1,2,...}\bar{m}_{n}E_{n}=E_{bp}N_0+\sum_{n\neq0}\bar{m}_{n}E_{n}
\end{equation}
With the use of (14), (15) and (19) the specific energy (i.e. the energy per one TI bipolaron) $\tilde{E}(\tilde{T})=E/N\omega^*$, $\tilde{E}_{bp}=E_{bp}/\omega^*$ will be:
\begin{eqnarray}\label{20}
    \tilde{E}(\tilde{T})=\tilde{E}_{bp}+\\\nonumber
		\frac{\tilde{T}^{5/2}}{C_{bp}}
		F_{3/2}\left(\frac{\tilde{\omega}-\tilde{\mu}}{\tilde{T}}\right)
		\left[\frac{\tilde{\omega}}{\tilde{T}}+
		\frac{F_{5/2}\left(\frac{\tilde{\omega}-\tilde{\mu}}{\tilde{T}}\right)}
		{F_{3/2}\left(\frac{\tilde{\omega}-\tilde{\mu}}{\tilde{T}}\right)}\right],
\end{eqnarray}
$$F_{5/2}(\alpha)=\frac{2}{\sqrt{\pi}}\int^{\infty}_0\frac{x^{3/2}dx}{e^{x+\alpha}-1},$$
where $\tilde{\mu}$ is determined by the equation:
\begin{equation}\label{21}
    \tilde{T}^{3/2}F_{3/2}\left(\frac{\tilde{\omega}-\tilde{\mu}}{\tilde{T}}\right)=C_{bp}
\end{equation}
$$\tilde{\mu}=\left\{\begin{array}{rl}
		0,\ \ \ \tilde{T}\leq\tilde{T}_c;\\
		\tilde{\mu}(\tilde{T}),\ \ \ \tilde{T}\geq\tilde{T}_c
		\end{array}\right. \ $$
Relation of $\tilde{\mu}$ with the chemical potential of the system $\mu$ is written as:
$\tilde{\mu}=(\mu-E_{bp})/\omega^*$. From (20)-(21) we can also get
the expressions for the free energy:  $\Delta F = - \frac{2}{3} \Delta E, \Delta F = F - E_{bp}N,  \Delta E = E - E_{bp}N$ and entropy $S=-\partial F/ \partial T$.

\begin{figure}
\begin{center}
\includegraphics[width=0.8\textwidth]{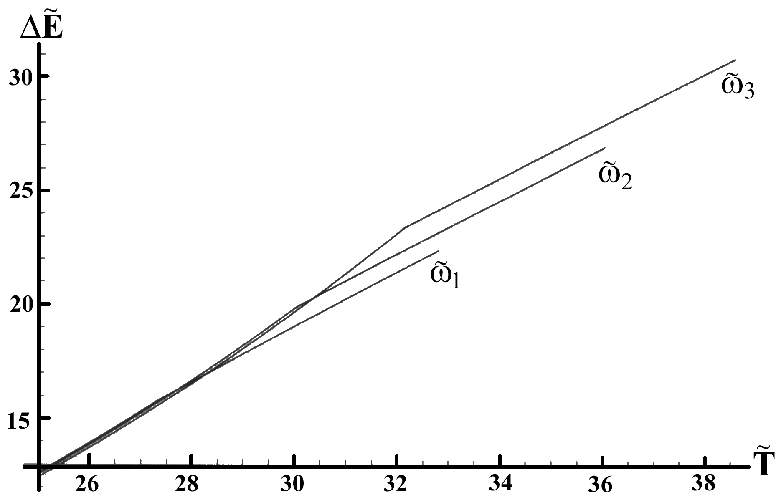}
\\
\includegraphics[width=0.8\textwidth]{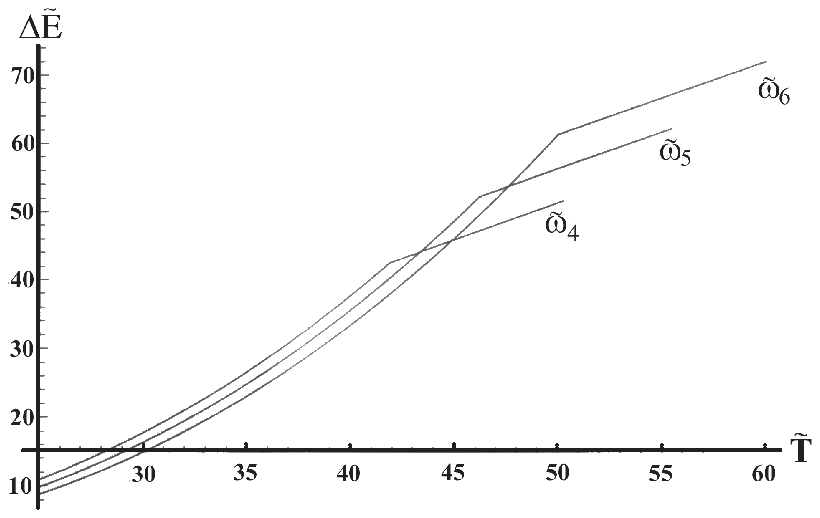}
\end{center}
\caption{Temperature dependencies $\Delta E(\tilde{T})=\tilde{E}(\tilde{T})-\tilde{E}_{bp}$
for the parameter values $\tilde{\omega}_i$ presented in Fig.~\ref{fig1}, \ref{fig2}.
\label{fig3}}
\end{figure}

Fig.~\ref{fig3} illustrates the temperature dependencies of $\Delta \tilde{E} =\tilde{E}-\tilde{E}_{bp}$
for the above-listed parameter values $\omega_i$.
The salient points on the curves $\Delta \tilde{E}_i(\tilde{T})$ correspond to the values of critical temperatures $T_{c_i}$.

\noindent The dependencies obtained enable us to find the heat capacity of a TI bipolaron gas: $C_v(\tilde{T})=d\tilde{E}/d\tilde{T}$.
With the use of (20) $C_v(\tilde{T})$ for $\tilde{T}\leq\tilde{T}_c$ is expressed as:

\begin{equation}
\label{22}
C_v(\tilde{T})=\frac{\tilde{T}^{3/2}}{2C_{bp}}
\left[\frac{\tilde{\omega}^2}{\tilde{T}^2}F_{1/2}\left(\frac{\tilde{\omega}}{\tilde{T}}\right)+
6\left(\frac{\tilde{\omega}}{\tilde{T}}\right)F_{3/2}\left(\frac{\tilde{\omega}}{\tilde{T}}\right)+
5F_{5/2}\left(\frac{\tilde{\omega}}{\tilde{T}}\right)\right],
\end{equation}
$$F_{1/2}(\alpha)=\frac{2}{\sqrt{\pi}}\int^{\infty}_0\frac{1}{\sqrt{x}}\ \frac{dx}{e^{x+\alpha}-1}$$

Expression (22) yields a well-known exponential dependence of the heat capacity
at low temperatures $C_v\backsim\text{exp}(-\omega_0/T)$,
caused by the availability of the energy gap $\omega_0$.

\begin{figure}
\begin{center}
\includegraphics[scale=0.8]{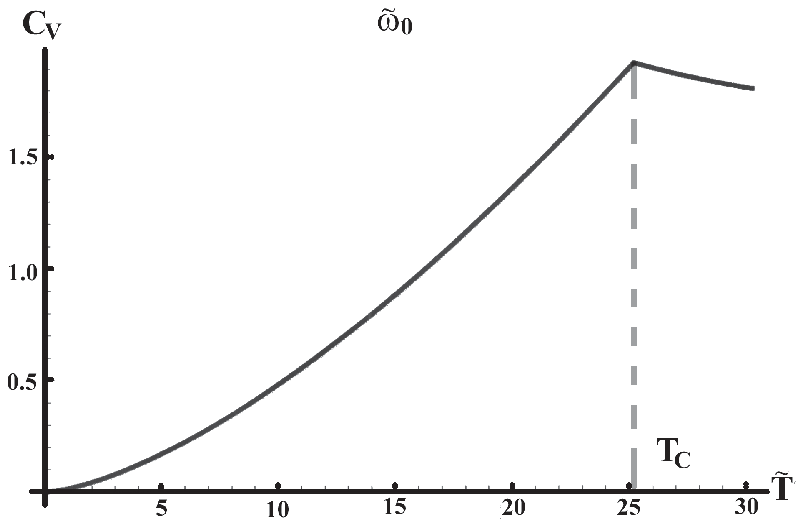}
\includegraphics[scale=0.9]{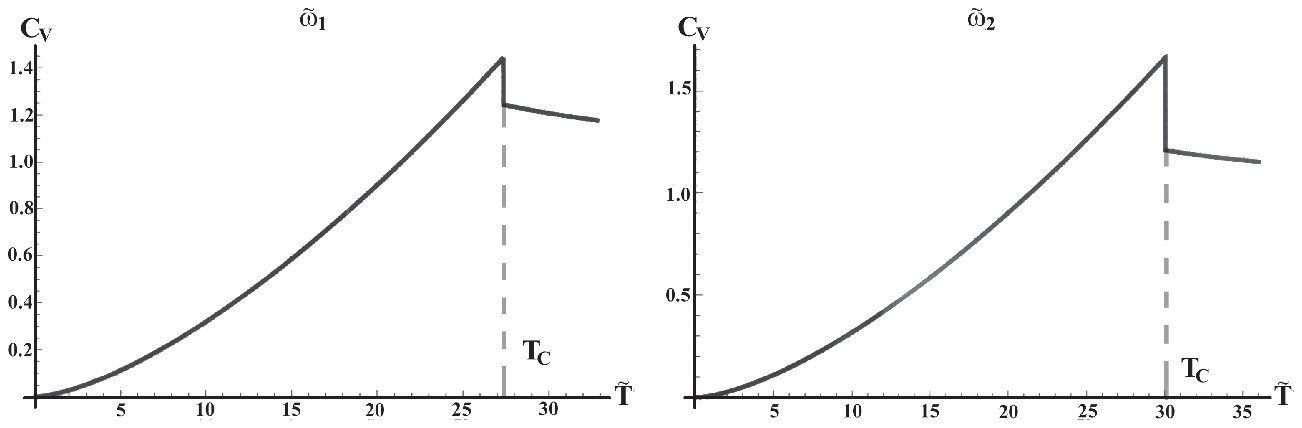}
\includegraphics[scale=0.9]{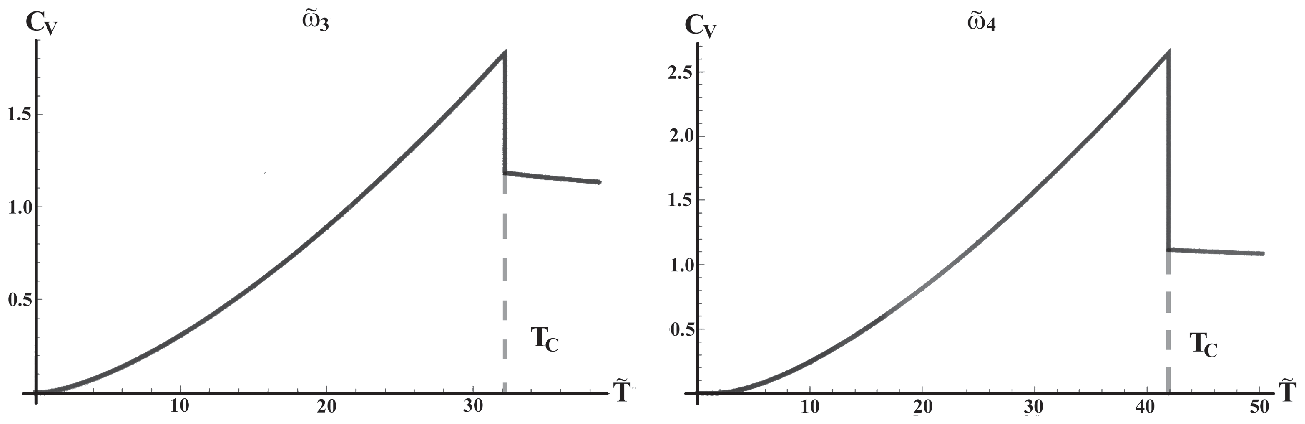}
\includegraphics[scale=0.9]{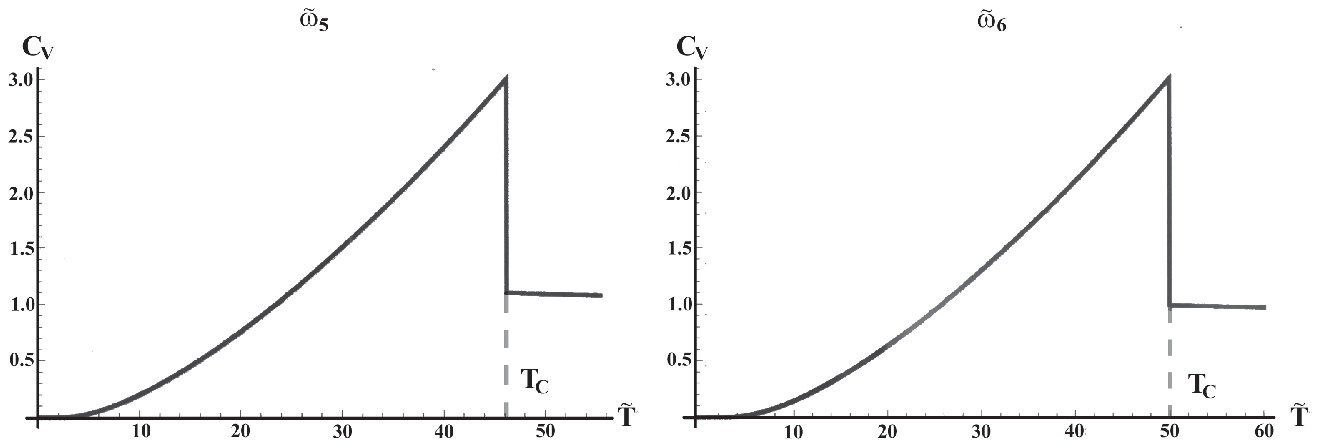}
\end{center}
\caption{
Temperature dependencies of the heat capacity for various
values of the parameters $\omega_i$: $\omega_0=0$;   $\ \tilde{T}_{C_0}=25,2$; $\ C_v(\tilde{T}_{c1})=2$;
$\omega_1=0,2$; $\ \tilde{T}_{C_1}=27,3$;  $\ C_v(\tilde{T}_{c1}-0)=2,16$; $\ C_v(\tilde{T}_{c1}+0)=1,9$;
$\omega_2=1$;   $\ \ \tilde{T}_{C_2}=30$; $\ C_v(\tilde{T}_{c2}-0)=2,46$; $\ C_v(\tilde{T}_{c2}+0)=1,8$;
$\omega_3=2$;   $\ \ \tilde{T}_{C_3}=32,1$; $\ C_v(\tilde{T}_{c3}-0)=2,7$; $\ C_v(\tilde{T}_{c3}+0)=1,78$;
$\omega_4=10$;  $\ \tilde{T}_{C_4}=41,9$; $\ C_v(\tilde{T}_{c4}-0)=3,7$; $\ C_v(\tilde{T}_{c4}+0)=1,7$;
$\omega_5=15$;  $\ \tilde{T}_{C_5}=46,2$; $\ C_v(\tilde{T}_{c5}-0)=4,2$; $\ C_v(\tilde{T}_{c5}+0)=1,65$;
$\omega_6=20$;  $\ \tilde{T}_{C_6}=50$; $\ C_v(\tilde{T}_{c6}-0)=4,6$; $\ C_v(\tilde{T}_{c6}+0)=1,6$.
\label{fig4}}
\end{figure}

Fig.~\ref{fig4} shows the temperature dependencies of the heat capacity $C_v(\tilde{T})$
for the above-listed values of the parameters $\tilde{\omega}_i$.
Table~\ref{tab1} lists the values of jumps in the heat capacity $\tilde{\omega}_i$ for the same parameter values:

\begin{equation}\label{23}
    \Delta \frac{\partial C_v(\tilde{T})}{\partial\tilde{T}}=
		\left. \frac{\partial C_v(\tilde{T})}{\partial\tilde{T}}\right|_{\tilde{T}=\tilde{T}_c+0}-
		\left. \frac{\partial C_v(\tilde{T})}{\partial\tilde{T}}\right|_{\tilde{T}=\tilde{T}_c-0}
\end{equation}
at the transition points.

The dependencies obtained enable us to find the latent heat of the transition $q=TS$,
where $S$ is the entropy of the supracondensate particles.
At the point of transition this value is equal to: $q=2T_cC_v(T_c-0)/3$,
where $C_v(T)$ is determined by formula (22) and for the above-listed values of $\omega_i$ is given in Table~\ref{tab1}.

The results obtained can be generalized to the case of a nonideal charged Bose gas if we replace $E_{bp}$ by $E_{bp}+u_0$ and $M$ by $M\sqrt{1+\chi}$ in formulas of this section. The results can also be generalized to the case when the dispersion of $\omega_0(k)$ takes the form $\omega_0(k)=\omega_0(0)+\beta k^2$. In this instance $\omega_0$ is replaced by $\omega_0(0)$ and $1/(2M_e)$ is replaced by $\beta + 1/(2M_e)$.

\section{Comparison with the experiment}

Success of the BCS theory is related to the fact that it has managed to explain
some experiments in ordinary metal superconductors where EPI is not strong.
There are grounds to believe that EPI in high-temperature ceramic superconductors
 is rather strong \cite{24}-\cite{26} and the BCS theory is hardly applicable to them.
In this case the bipolaron theory can suit. As is known, Eliashberg theory \cite{27}
which was developed especially to describe superconductors with strong
EPI fails to describe bipolaron states \cite{4}, \cite{5}.

Let us cite some experiments on HTSC which are in agreement with the TI-bipolaron theory.

According to the main SC theories available thus far (BCS, RVB, t-J theories \cite{9}, \cite{13}), at low temperatures all the current carriers should be paired (i.e. the density of superconducting electrons coincides with superfluid density). In recent experiments on overdoped SC \cite{28n} it has been shown that this is not the case: only a small part of current carriers appeared to be paired. Analysis of this situation made in \cite{29n} shows that the results obtained in \cite{28n} do not fit in with the theoretical constructions available. The TI-bipolaron theory of SC developed in this paper gives an answer to the question of the paper \cite{29n}, namely where most of the electrons disappear in the superconductors analyzed. It lies in the fact that only a small part of electrons $n_{bp}$: $n_{bp} \ll nE_{bp}/\epsilon _F \ll n$ occuring near the Fermi surface are paired and determine the superconducting properties of HTSC materials.

In fact, the theory of strong EPI developed here is not applicable to overdoped SC where weak  EPI coupling is expected and can hardly be used for explaining experiments on overdoped samples used in \cite{28n}. Particularly, in the underdoped samples we cannot expect a linear dependence of the critical temperature $T_c$ on the density of SC electrons observed in \cite{28n}. Rather this dependence would be nonlinear as it follows from equation (15).
For the overdoped regime, recently a theory \cite{Shaginyan} has been developed on the basis of the Fermi condensation idea \cite{Dukelsky}, which is a generalization of the BCS theory where the number of SC carriers was shown to be only a small part of their total number and which is in agreement with other observations of \cite{28n}.
Thus we can conclude that the phenomenon obtained in \cite{28n} is rather general and takes place both in overdoped and underdoped regimes (see also \cite{Bozovic2}).
We can also expect a linear dependence of the resistivity on $T$ if $T>T_c$ both in the overdoped and underdoped regimes since the number of bipolarons is small as compared with the total number of electrons and if EPI dominates in the  homogeneous crystal.

On the contrary to \cite{Shaginyan} it was shown in recent work \cite{Pashitskii} that the linear dependence of $T_c$ on the number of  Cooper pairs observed in \cite{28n} in the overdoped $La_{2-x}Sr_xCu_2O$ crystals can be described by BCS model for plasmon mechanism of SC. It seems nevertheless that the special case considered in \cite{Shaginyan} can not explain the general character of results obtained in \cite{28n}.

The problem of inconsistency of BCS with \cite{28n} was also considered in recent work \cite{Peeters-2018} where a simple bipolaron SC model was introduced and was shown that the number of bipolarons should be much less then the total number of carriers. The result obtained in \cite{Peeters-2018} confirms our results that only a small part of carriers are paired in the limit of low temperatures.

Fig.~\ref{fig3} shows typical dependencies of $E(\tilde{T})$. They suggest that at the point
of transition the energy is a continuous function of $E(\tilde{T})$.
This means that the transition per se occurs without energy expenditure being
a phase transition of the 2-kind in complete agreement with the experiment.
At the same time transition of Bose particles from a condensate state to
a supracondensate one occurs with consumption of energy which is
determined by the value of $q$ (\S3, Table~\ref{tab1}), determining the latent
heat of the transition of a Bose gas which makes it a phase transition of the 1-st kind.

By way of example let us consider HTSC $YBa_2Cu_3O_7$ (YBCO) with the temperature
of transition $90\div93$K, volume of the unit cell $0,1734\cdot10^{-21}$cm$^{-3}$,
concentration of holes $n\approx10^{21}$cm$^{-3}$.
According to estimates \cite{28}, the Fermi energy is equal to $\epsilon_F=0,37$ eV.
The concentration of TI-bipolarons in  $YBa_2Cu_3O_7$ is found from equation (15):
\begin{eqnarray}\label{24}
\frac{n_{bp}}{n}C_{bp}=f_{\tilde{\omega}}\left(\tilde{T}_c\right)
\end{eqnarray}
with $\tilde{T}_c=1,6$. Table~\ref{tab1} lists the values of $n_{bp,i}$ for the values
of parameters $\tilde{\omega}_i$  given in \S2. It follows from Table~\ref{tab1} that $n_{bp,i}<<n$.
Hence, only a small part of charge carriers is in a bipolaron state.
It follows that in complete agreement with the results of the previous section,
the Coulomb interaction will be screened by nonpaired electrons which justifies
the approximation of a noninteracting TI-bipolaron gas used by us.

According to our approach, superconductivity arises when coupled states are formed.
The condition for the formation of such states near the Fermi surface, by \cite{20},
has the form: $E_{bp}<0$. The value of the pseudogap, according to \S2, will be:
\begin{eqnarray}\label{25}
\Delta_1=\left|E_{bp}+u_0\right|
\end{eqnarray}
Naturally, this value is independent of the vector $\vec{k}$,
but depends on the concentration of current carriers, i.e. the level of doping.

In the simplest variant of the superconductivity theory presented here, the gap $\omega_0$
 does not change as the system passes on from the condensed state to the uncondensed one,
i.e. from the superconducting state to the nonsuperconducting one, therefore
$\omega_0$ has also the meaning of a pseudogap:
\begin{eqnarray}\label{26}
\Delta_2=\omega_0(\vec{k})
\end{eqnarray}
which depends on the wave vector $\vec{k}$.

Numerous discussions of the problem of a gap and a pseudogap rely on the statement
that the energy gap in HTSC is determined by the coupling energy of Cooper pairs
which leads to insoluble contradictions (see reviews \cite{29}-\cite{33}).

Actually, the value of the superconducting gap $\Delta_2$ (26) has no concern with
the energy of paired states which is determined by  $E_{bp}$.
As is shown in \cite{12}, the energy of a TI bipolaron is
$\left|E_{bp}\right|\backsim\alpha^2\omega_0$ for both small and large values of
$\alpha$ i.e. $\left|E_{bp}\right|$ does not depend on $\omega_0$ at all.

Thus, within our concept an answer to the question of why the pseudogap ($\Delta_2$)
has the same anisotropy as the superconducting gap is made clear -- this is one and
the same gap.  It also becomes evident why the gap and the pseudogap depend on temperature
only slightly. In particular, it becomes clear why under a superconducting transition
a gap arises immediately and does not vanish as $T=T_c$ (which is not typical for the BCS).
The oft-debated question of what order parameter should be put in correspondence
to the pseudogap phase (i.e. whether the pseudogap phase is a special state
of the matter \cite{30}) seems meaningless  within  the theory developed.

At present there are many methods for measuring the gap: angle-resolved
photoelectron spectroscopy (ARPES), Raman (combination) spectroscopy,
scanning tunnel spectroscopy, neutron magnetic scattering, etc.
According to \cite{33}, the maximum value of the gap $YBCO$ (6.6)
(in antinodal direction in ab-plane) was found to be  $\Delta_1/T_c\approx16$.
This gives $\left|E_{bp}\right|\approx 80$ meV.

\begin{figure}
\begin{center}
\includegraphics[width=0.9\textwidth]{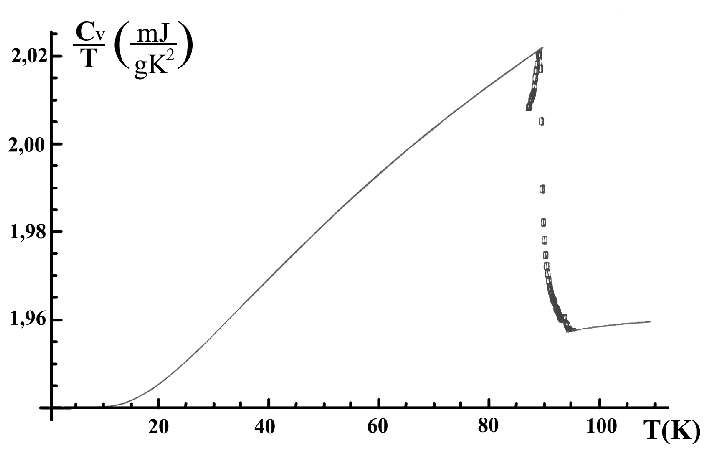}
\end{center}
\caption{Comparison of the theoretical (solid line) and experimental
(broken line) dependencies in the region of the heat capacity jump.
\label{fig5}}
\end{figure}

Now let us find a characteristic energy of phonons responsible for formation
of TI bipolarons and determining the superconducting properties of oxide ceramics,
i.e. the value of the superconducting gap $\Delta_2$. For this purpose we
will compare the calculated values of the heat capacity jumps with the experimental data (Fig.~\ref{fig5}).

\noindent As is evident from Fig.~\ref{fig5}, the theoretically calculated jump in the heat capacity (\S3)
coincides with the experimental values in $YBa_2Cu_3O_7$ \cite{34}, for $\tilde{\omega}=1,5$
i.e. for $\omega=7,5$ meV. This corresponds to the concentration
of TI bipolarons equal to $n_{bp}=2,6\cdot10^{18}$cm$^{-3}$. Taking into account that
$\left|E_{bp}\right|\approx0.44\alpha^2\omega$ \cite{11}, $\left|E_{bp}\right|=80$meV, $\omega=7,5$meV
the EPI constant $\alpha$ will be: $\alpha\approx5$,
and lies beyond the range of applicability of the BCS theory.

The availability of a gap $\omega_0$ in HTSC ceramics is proved by numerous
spectroscopy experiments (ARPES) on angular dependence of  $\omega_0$  on $\vec{k}$
 for small $|\vec{k}|$ \cite{29}-\cite{33}. The availability of d-symmetry
in the angular dependence of $\omega_0(\vec{k})$ is probably concerned
with arising of a pseudogap and rearragement of Fermi system into
the system of Fermi arcs possessing d-symmetry. In experiments
on tunnel spectroscopy $\omega_0$ can manifest itself as an occurrence of a pseudogap
structure superimposed on a pseudogap $\Delta_1$($\Delta_1>>\omega_0$).
In optimally doped  $YBa_2Cu_3O_7$ and $Bi_2Sr_2CaCu_2O_8$ (BCCO)
such a structure was observed many times in the region of
$5\div 10$ meV  \cite{35}-\cite{37},
which coincides with the above-cited estimation of $\omega_0$.

Many experimenters measure the dependence of the value of a gap and pseudogap
on the level of doping $x$. Even early experiments on magnetic responsibility
and Knight shift demonstrated the availability of a pseudogap which arises
for  $T^*>T_c$. Numerous subsequent experiments revealed peculiarities
of $T-x$ phase diagram: $T^*$ increases while $T_c$ decreases
as doping grows smaller \cite{29}-\cite{33}. As is shown in  \cite{20},
this behavior can be explained by peculiarities of the existence of bipolarons
in a polaron gas  \cite{20}. In  \cite{20} mention is also made
of possibly general character of 1/8 anomaly in HTSC systems.

In conclusion it should be noted that a longstanding discussion of the nature
of the gap and pseudogap in HTSC materials is in many respects associated
with the problem of measurement, when different measurement methods
in fact measure quite different quantities, rather than similar ones.
In the case under consideration, ARPES measures $\omega_0(\vec{k})$,
 while tunnel spectroscopy measures $\left|E_{bp}\right|$.
In this field unfortunately there are a lot of unsolved problems
which offer a challenge for both the theory and the experiment.

\section{Discussion of results}

The above considered theory of TI-bipolaron superconductivity as well as
the BCS theory rely on Froehlich Hamiltonian. These theories, however,
have different domains of applicability.  In the BCS theory EPI
is considered to be weak, accordingly, the correlation length,
or the characteristic size of the pair is $l_{corr}>>n^{-1/3}_{bp}$.
The concentration of bipolarons (i.e. Cooper pairs) $n_{bp}$
in the BCS is very large and for $T=0$ coincides
with the concentration of charge carriers in metals.

In the above considered theory EPI is considered to be strong,
therefore the correlation length is $l_{corr}<<n^{-1/3}_{bp}$.
At the same time even at $T=0$ bipolarons are only a small part of charge carriers.
This situation is realized in oxide ceramics.
The notion of a pair in the TI-bipolaron theory under these conditions is well determined.
Thus, according to \cite{2}, the value of the correlation length for a TI bipolaron is
$l_{corr}=\hbar^2\tilde{\epsilon}x(\eta)/e^2M_e$, where $x(\eta) (\eta=\epsilon_{\infty}/\epsilon_0)$ varies within $6\div10$
in the area of stability of bipolaron states.
For the values of parameters corresponding to  $YBCO$,
it makes up $l_{corr}\approx10^{-7}$cm while $n_{bp}^{-1/3}\approx10^{-6}$ cm.
Thus: $n_{bp}\,l_{corr}^3 \sim 10^{-3} \ll 1$, that is, individual pairs practically do not overlap.

The results obtained suggest that in order to raise the critical temperature $T_c$
one should either decrease the effective mass of charge carriers or increase the phonon frequency $\omega_0$,
or else enhance the concentration of bipolarons $n_{bp}$.
Hence, the problem of raising $T_c$ is related to the search for crystals with optimal parameter values.
Notice that an increase in $\omega_0$ will not necessarily lead to an increase in $T_c$,
 since an increase in $\omega_0$ leads to a decrease in the EPI constant $\alpha$.

For small $\alpha$  we find ourselves in the field of applicability of
the BCS which yields small $T_c$. This situation is probably realized
in the anticorrelation dependence of $T_c$ on $\omega_0$ \cite{38}.

The most effective way to change the concentration of $n_{bp}$ is to change
the level of doping. In this case both the concentration of charge carriers
$n$ and the concentration of $n_{bp}$ alter. The dependence of $n_{bp}(n)$
can be rather complicated. In particular, an increase in $n$ does not necessarily lead to an increase in $n_{bp}$ \cite{20}.

To enhance the transition temperature one can apply external pressure.
Enhanced pressure leads to a decrease in the crystal volume and,
accordingly an increase in the concentration of  $n_{bp}$,
a decrease in the effective mass of charge carriers  and a rise in the phonon frequency $\omega_0$.

An effective way to raise $T_c$ can be the use of unhomogeneous doping.
Thus, in a wire with cylindrical symmetry, doping as a result of which
the concentration of acceptors is maximal on the cylinder axis,
will lead to an increased concentration of bipolarons along the crystal's axis since
a Bose gas will concentrate in the regions with minimal potential energy.

Such conclusions do not account at all for many important factors.
Consideration of EPI alone is not adequate since in HTSC materials
of importance is the availability of a magnetic order.
Thus, for example, in many papers the occurrence of a pseudogap
is associated not with EPI, but with magnetic fluctuations \cite{39}.
Consideration of other interactions such as electron-plasmon interaction can also compete with EPI.

The theory of TI-bipolaron superconductivity as well as the BCS is based on isotropic 3D model.
Actually most cuprates are layered structures possessing high anisotropy.
Generalization of the theory to this case seems to be rather actual.
The situation is complicated by the fact that in real HTSC materials
of great importance are imperfections such as stripes and clusters.
In particular, the availability of stripes suggests a 1D scenario
of superconductivity in oxide ceramics.  TI-bipolaron mechanism leads
in this case to possible existence of a Bose condensate in 1D systems and,
as a consequence, to new opportunities – obtaining of HTSC in materials with long stripes \cite{40}.

Let us specify the main conclusions emerging from consistent translation-invariant consideration of EPI.
Electron pairing (for any coupling constant) based on such consideration
leads to a concept of TI polarons and TI bipolarons. Being bosons,
TI bipolarons can experience Bose condensation leading to superconductivity.
Let us point out the main consequences of this approach. First of all,
the theory under consideration resolves the problem of the great value
of the bipolaron effective mass (\S3). As a consequence,
formal restrictions on the value of the critical temperature
of the transition are removed.  The theory quantitatively explains
some thermodynamic properties of HTSC such as the availability (\S3)
and value (\S4) of the jump in the heat capacity which is lacking
in the theory of Bose condensation of an ideal gas.
It accounts for a large ratio of the pseudogap width to $T_c$ (\S4).
The small value of the correlation length \cite{2} is clarified.
The theory explains the occurrence of a gap and a pseudogap (\S3, \S4)
in HTSC materials. The angular dependence of a gap and pseudogap (\S4) is also clarified.

Accordingly, an isotropic effect straightforwardly follows from expression (13)
where phonon frequency $\omega_0$ plays the role of a gap.
Application of the theory to 1D and 2D systems yields qualitatively
new results since the availability of a gap in the TI bipolaron spectrum
eliminates divergencies which are observed for small impulses
in the theory of an ideal Bose gas \cite{40}.

\section*{Funding} The work was done with the support from Russian Science Foundation (RSF project N16-11-10163) and the Russian Foundation for Basic Research, (RFBR project N16-07-00305).

\section*{Appendix. Remarks on the notation}

Function $F_{3/2}(\alpha)$ is called a polylogarithm $=Li_{3/2}(e^{-\alpha})$,
in mathematics this is the function $PolyLog$, therefore the function $f_{\tilde{\omega}}$ in (15) will be:
$f_{\tilde{\omega}}=\tilde{T}^{3/2}PolyLog\left[3/2, e^{-\tilde{\omega}/\tilde{T}}\right]$.

In the general case, function $PolyLog$ of the order of $s$ is determined as:
$$PolyLog\left[s,e^{-\alpha}\right]=\frac{1}{\Gamma(s)}\int^{\infty}_0\frac{t^{s-1}}{e^{t+\alpha}-1}dt$$
where $\Gamma(s)$ is a gamma function: $\Gamma(1/2)=\sqrt{\pi}$, $\Gamma(3/2)=\sqrt{\pi}/2$, $\Gamma(5/2)=3\sqrt{\pi}/4$.

Accordingly, the functions $F_{1/2}$, $F_{3/2}$, $F_{5/2}$ occurring in the text will be:

$F_{1/2}=2PolyLog\left[1/2, e^{-\tilde{\omega}/\tilde{T}}\right]$;

$F_{3/2}=PolyLog\left[3/2, e^{-\tilde{\omega}/\tilde{T}}\right]$;

$F_{5/2}={3/2}PolyLog\left[5/2, e^{-\tilde{\omega}/\tilde{T}}\right]$.


\end{document}